\documentclass[twocolumn,superscriptaddress,showpacs,preprintnumbers,amsmath,amssymb,prl]{revtex4}
\usepackage{graphicx}
\usepackage{bm}
\usepackage[dvips]{color}              
\newcommand{\caco}{$\rm Ca_3Co_2O_6$}

\newcommand\AF{antiferromagnetic}
\definecolor{magenta}{rgb}{.5,0,.5}

\begin{document}
\title{The nature of the magnetic order in $\rm \bf Ca_3Co_2O_6$}

\author{S.~Agrestini}
\affiliation{Department of Physics, University of Warwick, Coventry, CV4 7AL, UK}

\author{L.~C.~Chapon}
\affiliation{ISIS facility, Rutherford Appleton Laboratory, Chilton-Didcot OX11-0QX, UK}

\author{A.~Daoud-Aladine}
\affiliation{ISIS facility, Rutherford Appleton Laboratory, Chilton-Didcot OX11-0QX, UK}

\author{J.~Schefer}
\affiliation{Laboratory for Neutron Scattering, ETH Z\"{u}rich and Paul Scherrer Institute, 5232 Villigen, PSI, Switzerland}

\author{A.~Gukasov}
\affiliation{Laboratoire L\'{e}on Brillouin (CEA-CNRS), CEA Saclay, 91191 Gif sur Yvette, France }

\author{C.~Mazzoli}
\affiliation{ European
Synchrotron Radiation Facility, BP 220, 38043 Grenoble Cedex 9, France}

\author{M.~R.~Lees}
\affiliation{Department of Physics, University of Warwick, Coventry, CV4 7AL, UK}

\author{O.~A.~Petrenko}
\affiliation{Department of Physics, University of Warwick, Coventry, CV4 7AL, UK} 

\date{\today}

\begin{abstract}
We present a detailed powder and single crystal neutron diffraction study of the spin chain compound \caco. Below 25~K, the system orders magnetically with a modulated partially disordered antiferromagnetic structure. We give a description of the magnetic interactions in the system which is consistent with this magnetic structure. Our study also reveals that the long-range magnetic order co-exists with a shorter range order with a correlation length scale of $\sim180$~\AA\ in the $ab$ plane. Remarkably, on cooling, the volume of material exhibiting short range order increases at the expense of the long-range order. 
\end{abstract}

\maketitle
The appearance of plateaux in the magnetization curves of low dimensional quantum spin systems (e.g. NH$_{4}$CuCl$_{3}$~\cite{Shiramura98} and SrCu$_{2}$(BO$_{3})_{2}$~\cite{Kageyama99}) has generated considerable attention from both an experimental and a theoretical point of view. In this respect the spin chain system \caco\ is very interesting because at low temperatures several steps, equally spaced in magnetic field,  appear in the magnetization~\cite{Kageyama97,Maignan00,Hardy04}, a behavior reminiscent of quantum tunneling of magnetization in molecular magnets~\cite{Gatteschi03}. The origin of this intriguing phenomenon is still an open question~\cite{Kudasov06,Maignan04} and the magnetic properties of \caco\ have been intensely studied in the past decade using many techniques, including x-ray~\cite{Takubo05,Burnus06,Agrestini08} and neutron~\cite{Aasland97,Kageyama98,Petrenko05} scattering, NMR~\cite{Sampa04}, calorimetry~\cite{Hardy03}, and magnetometry~\cite{Kageyama97,Maignan00,Hardy04}.

\caco\ is a rare example of a material where ferromagnetic (FM) 1D Ising spin chains are coupled through a much weaker \AF\ (AF) exchange on a triangular lattice. The spin chains in \caco\ are made up of alternating face-sharing octahedral (CoI) and trigonal prismatic (CoII) CoO$_6$ polyhedra, running along the $c$ axis and arranged in a triangular lattice in the $ab$ plane \cite{Fjellvag96}. The different Co environments leave the Co$^{3+}$ ions on the CoI sites in a low-spin ($S$=0) state, and those on the CoII sites in the high-spin ($S$=2) state~\cite{Sampa04,Takubo05,Burnus06}. Crystalline electric fields also lead to a very strong anisotropy with the moments preferentially aligned along the $c$ axis~\cite{Kageyama97,Maignan00}. The zero-field magnetic structure of \caco\ has yet to be described unambiguously. Most theoretical descriptions of the magnetic structure center around the two stable configurations for 1D Ising chains coupled antiferromagnetically on a 2D triangular lattice, the ferrimagnetic structure  $(M,M,-M)$ and the partially disordered \AF\ (PDA)  structure $(M,-M,0)$~\cite{Mekata77} (the labels in parenthesis indicate the relative magnetizations on the triangular lattice). The step-like magnetization has been described using the Ising model on a triangular lattice \cite{Kudasov06}. There are, however, experimental data which suggest a more complex magnetic structure for \caco. A pronounced drop in the intensity of the magnetic peaks on cooling below 18~K in zero field has been observed in powder~\cite{Aasland97,Kageyama98} and single crystal diffraction studies~\cite{Petrenko05}. Recent resonant x-ray scattering (RXS) studies~\cite{Agrestini08} showed a small ($\sim0.01$~\AA$^{-1}$) incommensuration in the magnetic reflections associated with a long-period modulation of the magnetic structure along the $c$ axis.  

In this work we address three fundamental questions that are key to understanding the physics, including the step-like behavior of the magnetization at low $T$, of \caco. 1) What is the true magnetic ground state of this material? 2) What is the physical origin of the observed modulation in the magnetic structure?  3) What produces the low $T$ reduction in the magnetic intensity seen in previous RXS and neutron scattering studies? To this end, we have carried out a careful investigation of \caco\ using neutron diffraction measurements as a function of $T$. The results reported here reveal that the actual magnetic order is neither a \textit{simple} PDA nor ferrimagnetic, but corresponds to a longitudinal sinusoidally modulated structure with a very long periodicity. We present a description of the magnetic interactions in this system, that can explain the observed modulation and emphasizes the need to consider the 3D character of the magnetic coupling in \caco. Finally, our data reveal that the reduction in the magnetic intensity of the neutron diffraction peaks is due to a coexistence of long and short range magnetic order. The results presented below provide a new insight into the nature of magnetic order in this geometrically frustrated material and call for a comprehensive revision of the theoretical models used to describe the magnetic behavior in \caco.

Single crystals of \caco\ were grown in a KCO$_3$ flux using \caco\ powder synthesized via a solid state reaction~\cite{Kageyama97}. A single crystal $\rm 15 \times 2 \times 1 mm^3$ with the longest direction parallel to the $c$ axis was used for the neutron diffraction experiments. The high quality of the crystals was confirmed by x-ray diffraction, energy dispersive x-ray, magnetization, and specific heat measurements.
\begin{figure}[tb]
\begin{center}
\includegraphics[width=1.0\columnwidth]{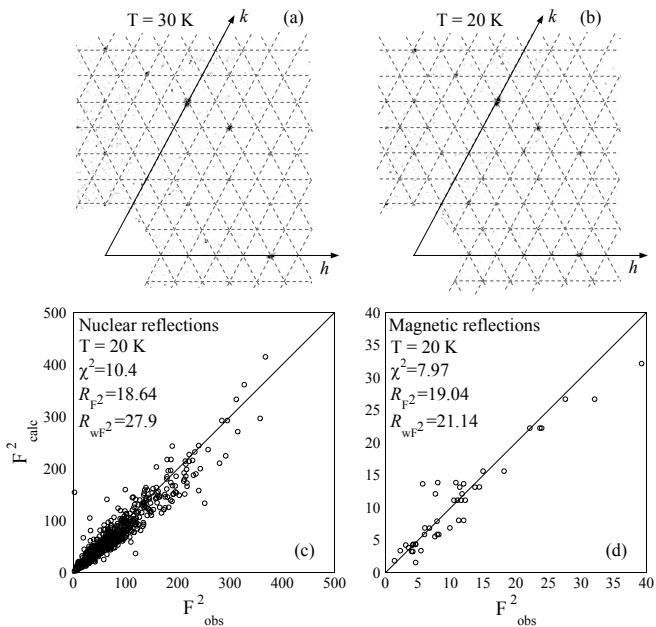}
\caption{\label{fig1} Single crystal neutron diffraction data for \caco\ taken on the time-of-flight SXD instrument at ISIS ~\cite{Gutmann06}. Neutron intensity patterns were recorded in the $(hk0)$ scattering plane at three different temperatures (a) $T$=30~K, (b) 20~K and 2.2~K (not shown), above and below the magnetic ordering temperature, $T_{N}$=25 K. A refinement of the data~\cite{SANote2} ($\chi^{2}$=9.18, $R_{F^{2}}$=18.84, $R_{wF^{2}}$=24.53) revealed that no structural transitions occur on cooling from 30~K down to 2~K; there are no significant variations in the lattice reflections or the atomic parameters, which are in good agreement with those in the literature~\cite{Aasland97,Fjellvag96}.
}
\end{center}
\end{figure}

Neutron single crystal diffraction experiments were performed on the SXD time-of-flight instrument at the ISIS-RAL, UK. SXD uses the white beam Laue technique, a stationary crystal, and a large area position-sensitive detectors covering a solid angle of $\sim2\pi$ sr~\cite{Keen06}, allowing a quick data collection over a large area of reciprocal space. Neutron powder diffraction experiments were conducted on the GEM instrument (ISIS) and were used to follow the $T$ evolution of the magnetic structure. The 4-circle TriCS thermal neutron single crystal diffractometer at the SINQ facility of the PSI, Switzerland was used for precise measurements as a function of $T$ of the position, intensity, and width, of a selection of magnetic and nuclear peaks. With a graphite (002) monochromator providing a wavelength of 2.32~\AA, the TriCS resolution $\Delta d/d$ was about 0.5\%.

\begin{figure}[tb]
\begin{center}
\includegraphics[width=1.0\columnwidth]{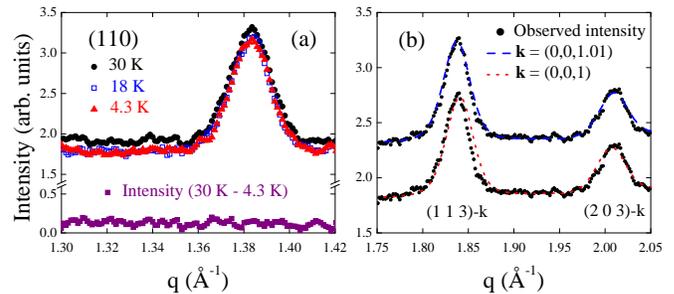}
\caption{\label{fig2}(color online). A selection of the results of the powder measurements on the time-of-flight GEM diffractometer at ISIS. (a) Neutron intensity of the $(110)$ nuclear Bragg peak in \caco\ at different temperatures; no variations in intensity are observed except for an increase of the background at $T$=30~K due to a paramagnetic contribution. (b) Neutron diffraction pattern at 18~K {\large($\bullet$)} and refinement shown for two values of the propagation vector (red dotted line, $k_{z}=1.00$) and (blue dashed line, $k_z=1.01$)~\cite{LCNote2}.
}
\end{center}
\end{figure}

The results of the SXD measurements are shown in Fig.~\ref{fig1}. Magnetic reflections appearing below T$_N$=25 K can be indexed by the propagation vector \textbf{k}=(0,0,1) with respect to the hexagonal setting of the rhombohedral space group R$\overline{3}c$ ~\cite{LCNote2}. The only apparent difference between the data collected at 20 and 2.2~K is an $\sim$20\% drop in the intensity of the magnetic peaks at low $T$, as previously reported~\cite{Aasland97,Kageyama98,Petrenko05}. All the $(hkl)$ magnetic peaks have even $l$, which supports the idea that the intra-chain ordering is ferromagnetic. The systematic absence of the $(00l)$ class of reflections confirms the Ising nature of the magnetic system with the Co magnetic moments aligned parallel to the $c$ axis. Previous neutron diffraction studies have suggested \caco\ adopts a ferrimagnetic structure~\cite{Aasland97,Kageyama98}. This assumption can be rejected as this would lead to additional magnetic contributions at \textbf{k}=0, and especially in the (110) and (300) peaks, that are not observed experimentally in either single crystal or powder neutron measurements (see Fig.~\ref{fig2}a). Our SXD single crystal data is consistent with a simple PDA structure. The refined value at $T$=20~K of the magnetic moment on the high-spin ($S$=2) CoII ion is 5.0 $\pm$ 0.1 \emph{$\mu$$_{B}$}~\cite{SANote1} and includes a sizeable orbital moment ($L$$\sim$1). Magnetic dichroism measurements on \caco\ gave similar values for the spin and orbital moments on the CoII ion~\cite{Burnus06}. 

\begin{figure}[tb]
\begin{center}
\includegraphics[width=0.85\columnwidth]{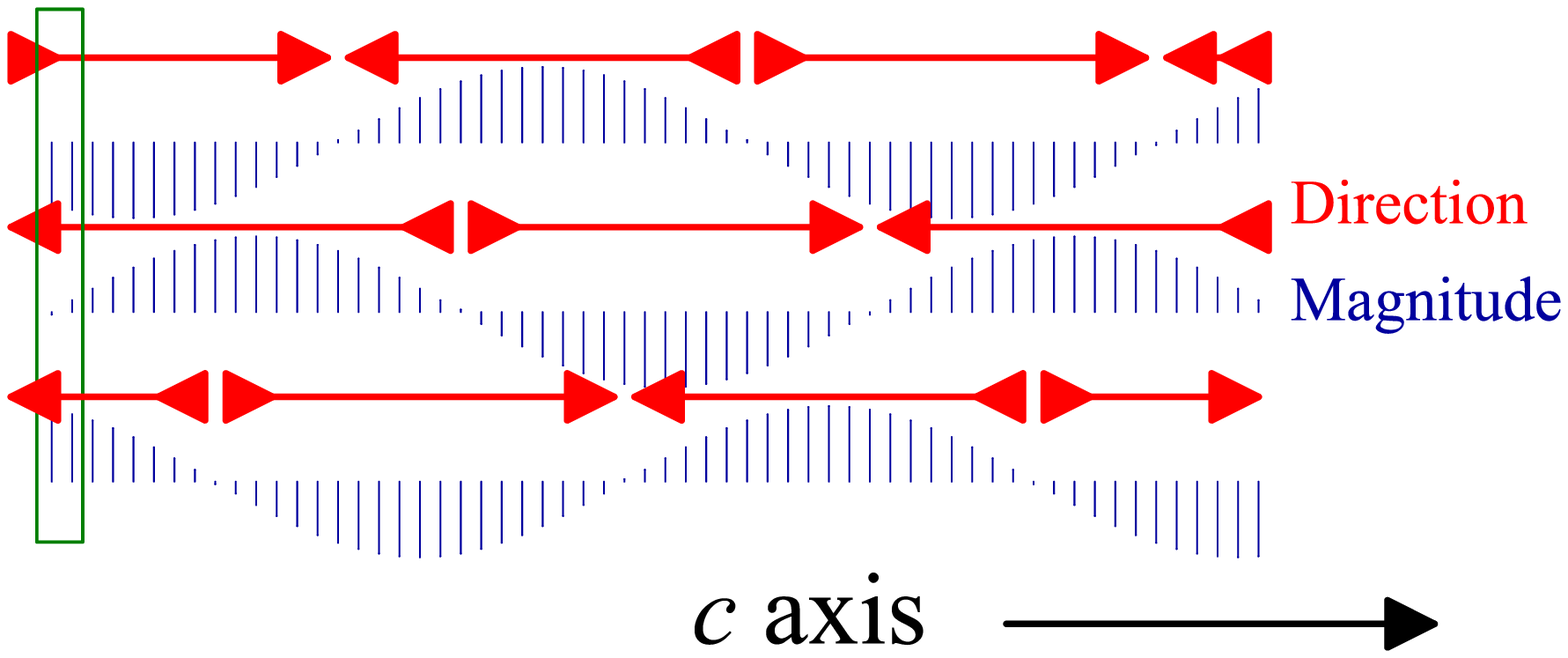}
\includegraphics[width=0.85\columnwidth]{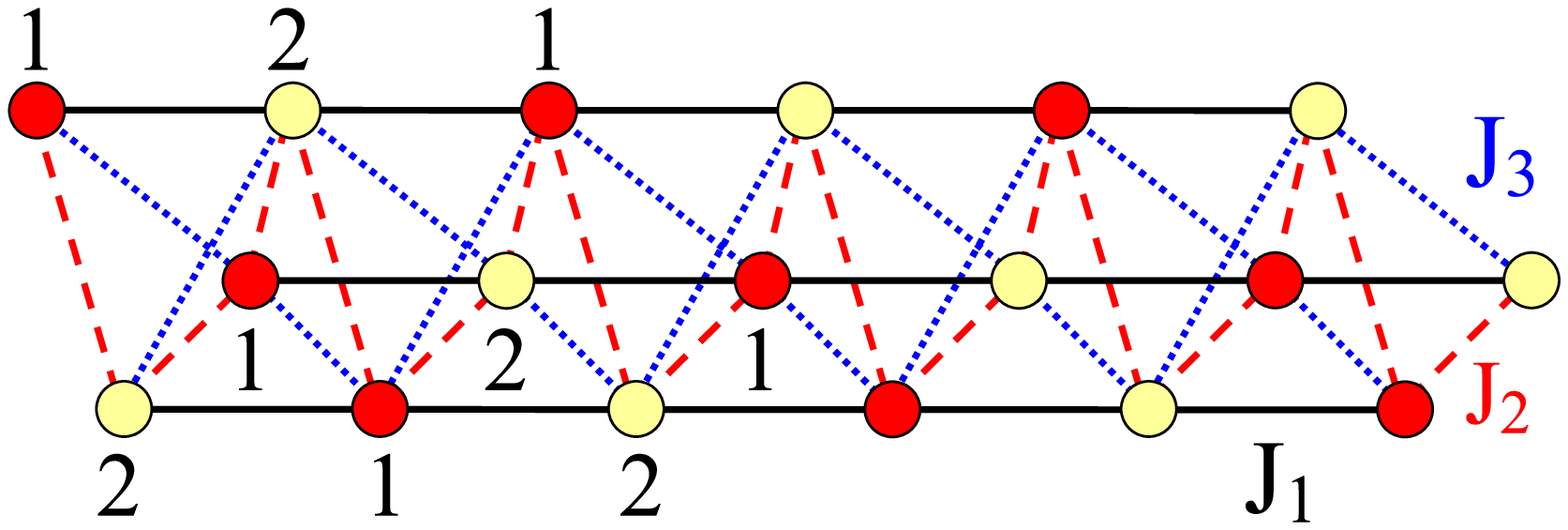}
\caption{\label{fig3}(color online). (a) Schematic of the proposed incommensurate magnetic structure for 30 unit cells along the $c$ axis with $k_z$ increased to 1.05 to emphasize the modulation of the moments. The magnitude and direction at sites in three adjacent chains are shown.  (b) Magnetic exchange interactions in \caco\: $J_{1}$ (black lines) FM; $J_{2}$ (red dashed lines) AF; $J_{3}$ (blue dotted lines) AF. CoII sites at (0,0,$\frac{1}{4}$) and (0,0,$\frac{3}{4}$) are labeled 1 and 2.
}
\end{center}
\end{figure}

GEM powder diffraction data, of slightly higher resolution, reveal that the magnetic propagation vector is in fact incommensurate with the crystal structure, \textbf{k}=(0,0,$k_z$) $k_z\sim$1.01, in excellent agreement with recent results obtained with magnetic x-ray scattering ~\cite{Agrestini08}. This is illustrated in Fig.~\ref{fig2}b, where a commensurate value for $k_z$ fails to reproduce the correct position of the magnetic Bragg peaks. Symmetry analysis using representation theory shows that the magnetic representation $\Gamma$ for a magnetic ion on site 6a $(0,0,\frac{1}{4})$ is decomposed into three irreducible representations: $\Gamma=\Gamma_{1}+\Gamma_{2}+2\Gamma_{3}$ ~\cite{LCNote1}. Only the symmetry-adapted mode belonging to $\Gamma_{1}$ allows a fit to the data. The magnetic phase transition, involving a single irreducible representation, is therefore consistent with the Landau theory of second-order transitions. For this mode the moments lie along the $c$ axis and their amplitude can be calculated using 
$\mathbf{M_1(R_L)}= M cos \left( 2\pi\mathbf{k \cdot R_L}\right)$ and $\mathbf{M_2(R_L)}=-M cos \left( 2\pi\left(\mathbf{k \cdot R_L} + \frac{k_z-1}{2}\right)\right)$ where the subscript 1 and 2 refer to symmetry related sites 1:$(0,0,\frac{1}{4} )$ and 2:$(0,0,\frac{3}{4} )$, and $\mathbf{R_{L}} = (R_{x},R_{y},R_{z})$ is the translation vector with respect to the zeroth-cell including R-centering translations. The resulting magnetic arrangement with an extended longitudinal modulation, ($\sim$1000~{\AA} or 200 CoII ions) is illustrated in Fig.~\ref{fig3}a. If the periodicity is truly incommensurate with the nuclear lattice, ($k_z$ is not a fractional value), any value of moment between -$M$ and +$M$ is found somewhere in the lattice. At specific lattice points, the magnetic configuration within nearest neighbor sites in adjacent chains of the triangular lattice is exactly the simple PDA structure with $(M,-M,0)$ configuration. At other lattice points, the configuration is exactly $(M,-M/2,-M/2)$. Other intermediate situations are also found. This type of structure is usually induced by competing exchange interactions in the presence of strong axial anisotropy, stabilizing an arrangement in which the ordered component of the magnetic moment fluctuates along the propagation direction.

\caco\ is often described in terms of Ising chains on a triangular lattice. Within this framework, the AF in-plane correlations and the FM correlations along the c axis are completely decoupled. Clearly, the magnetic energy of such a lattice is not lowered by introducing a small incommensurability, so the approximation to a 2D lattice must be revised. The Co ions in adjacent chains in \caco\ are shifted out of plane by $\frac{1}{6}$ or $\frac{2}{6}$ of $c$~\cite{Fjellvag96}. A sketch showing the interactions within a single triangular unit in the structure and omitting the oxygen atoms for simplicity, is shown in Fig.~\ref{fig3}b. Within each chain, direct Co-Co overlap dominates and leads to a strongly ferromagnetic interaction $J_{1}$~\cite{Fresard04}. The inter-chain coupling is more complex. There are two inequivalent AF super-superexchange interactions involving Co-O-O-Co paths labeled as $J_{2}$ and $J_{3}$, that follow helical paths and connect the Co sites in adjacent chains. For $J_{2}$, the overlap of the O 2p orbitals is very small making $J_{3}$ the dominant term after $J_{1}$~\cite{Fresard04}. This helical exchange pathway introduces an AF inter-chain coupling between the nearest neighbor CoII ions that competes with the intra-chain FM interaction producing the observed modulation along the $c$ axis. This type of competition leads naturally to non-colinear helicoidal arrangement for weak single-ion anisotropy, or a longitudinal sinusoidally modulated structure in the presence of large axial anisotropy as described here. We note that the present structure contains extended \emph{regions} with very weak static magnetic moments, a situation that favors the development of defects or dislocations along the chains. 
\begin{figure}[tb]
\begin{center}
\includegraphics[width=0.9\columnwidth]{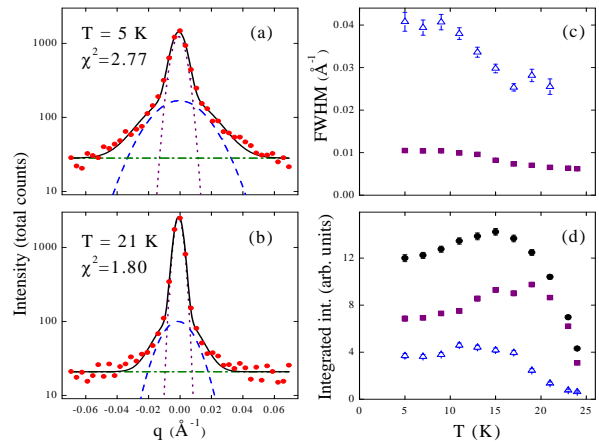}
\caption{\label{fig4} (color online). Transverse scans in the $ab$ plane through the $(100)$ magnetic
Bragg peak at (a) $T$=5~K and (b) $T$=21~K. The data were collected on the 4-circle TriCS diffractometer at PSI. Each peak is fitted (full line) by the sum of a broad and a narrow Gaussian function, together with a flat background. (c) FWHM and (d) integrated intensity of the broad (\textcolor{blue}{$\bigtriangleup$}) and narrow (\textcolor{magenta}{$\blacksquare$}) Gaussian components of the $(100)$ peak as a function of $T$. The total integrated intensity {\large($\bullet$)} is also shown.  
}
\end{center}
\end{figure}

Finally, we address the origin of the reduction in intensity of the magnetic peaks. Figures~\ref{fig4}a and ~\ref{fig4}b show transverse scans in the $ab$ plane through the $(100)$ magnetic Bragg peak as measured on the TriCS diffractometer at $T$=5~K and $T$=21~K. The peaks are fitted using two Gaussian functions with significantly different widths, the narrower of which corresponds closely to the instrumental resolution. The $T$ dependence of the peak widths are shown in Fig.~\ref{fig4}c. Above $21$~K, the peak cannot be fitted reliably with the two components, therefore the data are truncated at this point. At $T$=21~K, the ratio of the intensities of broad and narrow components is 0.2; this ratio increases to 0.7 at low $T$. A similar behavior has been observed for several other magnetic peaks, including the $(200)$, $(112)$ and $(102)$ and in difference plots of the GEM powder diffraction data collected at different temperatures. The loss of integrated intensity in the narrow component is mirrored by an increase in intensity of the broader feature as well as in the overall background. In other words, the anomalous reduction in the intensity of the magnetic peaks is due to the onset of a short-range magnetic ordering. From the reflection widths measured at $5$~K, the estimated correlation length for the short range order is $180$~\AA\ in the $ab$ plane~\cite{SANote3}. This is a clear sign of the increasing instability of the longer-range magnetic order as the temperature is reduced. One can speculate that the onset of shorter-range correlations originates from defects in the propagation of the long-range magnetic structure presented earlier, with stacking faults developing along the $c$ direction. These results are clearly different from the case of Ca$_{3}$CoRhO$_{6}$ where short range magnetic order is observed above and below the long range ordering temperature of $100$~K,  but there is little or no $T$ dependence to the diffuse component below $100$~K~\cite{Loewenhaupt03}.

In summary, neutron diffraction measurements show that the magnetic structure in \caco\ is a complex modulated PDA structure running along the $c$ axis. We propose a description of the magnetic interactions in terms of helical path couplings. This 3D description can explain the observed modulation along the chains. Our data reveal the appearance, at low $T$, of a short range magnetic structure that coexists with the long range order. We suggest that this phase coexistence is the origin of the drop in the intensity of the antiferromagnetic reflections at low $T$ observed in previous experiments.

This work was supported by a grant from the EPSRC, UK (EP/C000757/1) and by the European Commission under the 6th Framework Programme through the Key Action: Strengthening the European Research Area, Research Infrastructures. Contract n$^{o}$: RII3-CT-2004-506008 and RII3-CT-2003-505925.


\begin{thebibliography}{100}
\bibitem{SANote2} The crystal used for the SXD experiment was twinned. The twinning consists of a non-merohedral reverse/obverse twin. The data consisted of two components satisfying the reverse/obverse reflection condition ($h-k+l=3n$)/($-h+k+l=3n$) with an intensity ratio of 0.473(2). Fullprof ~\cite{Carvajal93} allowed us to use the data from both twin components in the structural refinement.
\bibitem{LCNote2} \caco\ is usually described in terms of a hexagonal cell but the system is actually trigonal. In the primitive unit cell, the magnetic propagation vector lies on a line of symmetry $(x, x, x)$ and is transposed to $(0, 0, 3x)$ in the hexagonal R setting. There is nothing special about $x=\frac{1}{3}$ which is inside the Brillouin zone and so the modulation does not split the magnetic peaks when $x\neq\frac{1}{3}$.
\bibitem{SANote1} In our refinement we have fixed at zero the magnetic moment of the CoI ion because it is now well established~\cite{Sampa04,Takubo05,Burnus06} that the CoI ion is in a LS spin state.
\bibitem{LCNote1}The notation follows Kovalev's conventions; $\Gamma_{1}$ and $\Gamma_{2}$ are 1D representations and $\Gamma_{3}$ is two-dimensional.
\bibitem{SANote3} The magnetic correlation length $D$(\AA) is estimated using $D$=2$\pi$/FWHM
\end{thebibliography}
\end{document}